\documentclass[preprint,12pt]{elsarticle}

\usepackage[english]{babel}		
\usepackage{mathptmx}					
\usepackage[bookmarks=true, colorlinks=true, urlcolor=blue, linkcolor=blue, citecolor=blue, citecolor=blue]{hyperref}							

\usepackage{tabularx}					

\newcommand{\bq}{\begin{quote}}
\newcommand{\eq}{\end{quote}}

\newcommand{\beq}{\begin{equation}}
\newcommand{\eeq}{\end{equation}}

\newcommand{\bfig}{\begin{figure}}
\newcommand{\efig}{\end{figure}}

\newcommand{\keff}[1]{\ensuremath{k_{eff}^{#1}}}
\newcommand{\ep}[1]{\ensuremath{\times 10^{#1}}}

\newcommand{\rhoPu}{\ensuremath{\rho_{Pu}}}

\newcommand{\ck}{\ensuremath{c_k}}
\newcommand{\corr}{\ensuremath{corr}}
\newcommand{\wtp}[1]{\ensuremath{wt\,\%\,^{#1}}}

\journal{Annals of Nuclear Energy}

\begin{document}
\begin{frontmatter}

\title{Analysis of experimental series of plutonium nitrate in aqueous solution and their correlation coefficients}

\author{R. Kilger, F. Sommer, M. Stuke}
\address{Gesellschaft f\"ur Anlagen- und Reaktorsicherheit (GRS) gGmbH,
\newline
Boltzmannstra{\ss}e 14, 85748 Garching, Germany}
\ead{Robert.Kilger@grs.de, Fabian.Sommer@grs.de, Maik.Stuke@grs.de}

\begin{abstract}
In this work we performed a detailed analysis on the calculation of 43 critical experiments from 6 experimental series all describing plutonium nitrate in aqueous solution contained in metal spheres. 
The underlying experimental data is taken from the handbook of the International Criticality Safety Benchmark Evaluation Project (ICSBEP) Working Group.
We present our modeling assumptions which were derived from the interpretation of the experimental data and discuss the resulting sensitivity analysis.
Although the experiments share some components, the derived correlation coefficients are for many cases statistically not significant.
Comparing our findings for the correlation coefficients with available data from the DICE Database we find an agreement for the correlation coefficients due to nuclear data. 
We also compare our results for the correlation coefficients due to experimental uncertainty. 
Our findings indicate that for the reliable determination of correlation coefficients a detailed study of the underlying experimental data, the modeling approach and assumptions, and the resulting sensitivity analysis seems to be inevitable.

\end{abstract}


\end{frontmatter}

\section{Introduction}
\label{sec:introduction}

The prediction of the effective neutron multiplication factor \keff{} below an appropriate safety margin is essential in criticality safety assessments.
The determination of \keff{} values and the resulting safety margins are performed by using validated calculation methods with validated computer codes, so called criticality codes, e.g. within the SCALE package \cite{ORNL.2012}, used in this work.
To validate a code, suitable critical benchmark experiments performed in various laboratories around the world are modeled and calculated. 
A large evaluated set of critical experiments is documented in the International Handbook of Evaluated Criticality Safety Benchmark Experiments (ICSBEP) \cite{NuclearScienceCommittee.September2014}. 
The experiments are mostly conducted in series, sharing some components, e.g. tanks, fuel rods or solutions. 
This can lead to significant correlations between experiments which in some cases have to be considered when determining safety margins \cite{Hoefer:2014xua, Sobes.2015, Peters.2015b, Peters.2016, Peters2016355}.
One possibility to quantify the degree of linearity between two quantities $X$ and $Y$ (e.g. two \keff{} values) is the Pearson correlation coefficient:
\beq
corr(X,Y) = \frac{cov(X, Y)}{\sigma(X)\sigma(Y)} 
\eeq
\beq
\quad \mathrm{with} \quad cov(X,Y) = (X-[X]) \times (Y-[Y])
\eeq
$\sigma(X)$ is the standard deviation and $[X]$ the expectation value of $X$. 
The Pearson correlation coefficient is defined in the interval $[-1,1]$ and does not account for non-linear relations between the two values.
Note, that TSUNAMI's \ck{} is defined the same way, using uncertainties in the nuclear data. 

In this paper we use the definition that the correlation between \keff{} and a varied parameter is called the sensitivity of \keff{} on the varied parameter.

The work presented in this paper was motivated by recent contributions of our group to a Benchmark defined by the Expert Group on Uncertainty Analysis for Criticality Safety Assessment (EGUACSA), a subgroup of the Working Party on Nuclear Criticality Safety (WPNCS) within the the Nuclear Energy Agency of the Organisation for Economic Co-operation and Development (OECD-NEA) \cite{Peters.2015b, Peters.2016, Peters2016355, Peters.2015}.
The Benchmark entitled "Role of Integral Experiment Covariance Data for Criticality Safety Validation" \cite{Hoefer.Oct.2014} discusses the influence of experimental uncertainties, covariances and correlations between critical benchmark experiments used for validation, bias and safety limits determination. 

While the Benchmark considered water moderated arrays of fuel rods, in this work we examine water reflected spheres of low concentrated plutonium nitrate solution with a thermal neutron spectrum.
In the ICSBEP these experiments have the identifier PU-SOL-THERM (PST). 
Different sizes of spheres, different \wtp{240}Pu and different plutonium nitrate concentrations are considered.
One series consists of one size of spheres and several experiments have the same plutonium content of $^{240}$Pu. 
Therefore it is plausible that these experiments are not statistically independent data sets and that neglecting correlations in a validation process could lead to incorrect results.

In the following work we analyze the experiment series PST-03 to -06, and -20 and -21 and determine their correlation matrix.
The paper is structured as follows: Section \ref{sec:experiments} describes the experimental setup, section \ref{sec:modelingAssumptions} describes the modeling assumptions. 
Section \ref{sec:results} shows the results of the calculation: section \ref{secSub:keff} shows the \keff{} values including nominal values and all uncertainty analysis, \ref{secSub:uncAnalysis} the uncertainty analysis due to system parameters, section \ref{secSub:corrSystemParams} the correlations due to uncertainties of system parameters, section \ref{secSub:corrNuclearUnc} the correlations due to nuclear data, and section \ref{secSub:dice} compares our findings with the data given in the DICE databank \cite {DICE}.
In section \ref{sec:discussionConclusion} we discuss our results and conclude in section \ref{sec:Conclusion}.

\section{Experiments}
\label{sec:experiments}

A total of 43 experiments from 6 experimental series (PST-03 to -06, and -20 and -21) was analyzed. 
All experiments describe plutonium nitrate in aqueous solution, contained in metal spheres. 
All experiments have a thermal spectrum and are slightly under-moderated.
Series PST-03, 04, 05 and 06 have some dissolved iron as impurity in the solution.
A list of all experiments can be found in table \ref{tab:experiments} showing the 6 calculated series, the considered experiments, the diameter of the spheres in inch, the experimental uncertainty, and if the latter was calculated by the evaluator or assumed from similar experiments.
\begin{table}[ht]
	\centering
		\caption{Analyzed experiments.}
		\label{tab:experiments}
		\begin{tabular}{|l|l|c|l|c|l|}
			\hline
			Experimental & Exp. & \# Exp. & Color in & Diameter   & Exp. \\
			series       &  		& 				& figures  &of sphere  & uncertainty \\
			\hline
			PST-03	& 01-08 &  8 & black 	& 13''   & 0,0047 assumed worst case \\
			\hline                 
			PST-04	& 01-13 & 13 & red 		& 14''   & 0,0047 assumed worst case \\ 
			\hline                 
			PST-05	& 01-09 &  9 & green	& 14''   & 0,0047 assumed worst case \\
			\hline                 
			PST-06	& 01-03 &  3 & blue		& 15''   & 0,0035 calculated \\
			\hline                 
			PST-20	& 10-15 &  6 & purple	&	14''   & 0,0047 calculated \\
			\hline                 
			PST-21	& 07-10 &  4 & cyan		& 15.2'' & 0,0025-0,0044	calculated \\
			\hline
		\end{tabular}
\end{table}

Almost all spheres consist of stainless steel. 
Only experiments PST-03-07 and 08 use an aluminum sphere. For series PST-21 a simplified model is assumed, which does not comprise the metal sphere and is corrected to account for the implications of this modification. 
All spheres were almost completely filled with solution when reaching criticality. 
The remaining void was compensated in the experiment description by adjusting the density of the solution. 
The same was done for the solution in the inlet, the outlet and the instrumentation pipes, the structural components, and the measuring equipment. 
Experimentally, almost all spheres are submerged in a rectangular water tank with at least 30~cm of surrounding water. 
Since from a modeling perspective 30~cm of water reflector is equivalent to an infinite water reflector, a spherical approximation of this water is a valid approximation. 
Only experiments PST-07, 08 and 09 from series 21 are bare spheres without any reflector. 

These simplifications and the accompanying compensations allow a very simple, spherical symmetrical computational model of the experiments: the spherical metal tank with the homogeneous plutonium nitrate solution and a surrounding water sphere of 30~cm. 
Experiments 14 and 15 of series PST-20 have an additional cadmium coating of 0.03 inch (0.762~mm) on the sphere. 
The detailed descriptions of the experiments in reference \cite{NuclearScienceCommittee.September2014} include all assumptions and simplifications.

\section{Modeling assumptions}
\label{sec:modelingAssumptions}

The experiments are modeled with the criticality code sequence CSAS5 of the code packet SCALE 6.1.2 \cite{ORNL.2012}. 
The continuous energy cross-section library ce\_v7\_endf based on the ENDF/B-VII library is used for the CSAS5 calculations.
The impact of nuclear data uncertainties on \keff{} was analyzed with TSUNAMI of the same code package.
For these calculations the cross-section library v7-238 with a 238 energy-group-structure was employed, using CENTRM for the resonance self-shielding calculations.

In the CSAS5 calculations 10.000 neutrons are followed, the first 100 generations are skipped and the calculation is stopped, when the Monte Carlo precision drops below 1\ep{5}. 
This value was typically reached after a total of 430 generations.

For the nominal cases, the mass number densities of the solutions, spheres and the surrounding water are taken directly form the experimental description. 
For plutonium they are given for all isotopes; for nitrogen, hydrogen, oxygen, iron, chrome, nickel, manganese, aluminum and cadmium the natural abundances are used.
\begin{figure}[ht]
    \centering
    \includegraphics[width=0.9\textwidth]{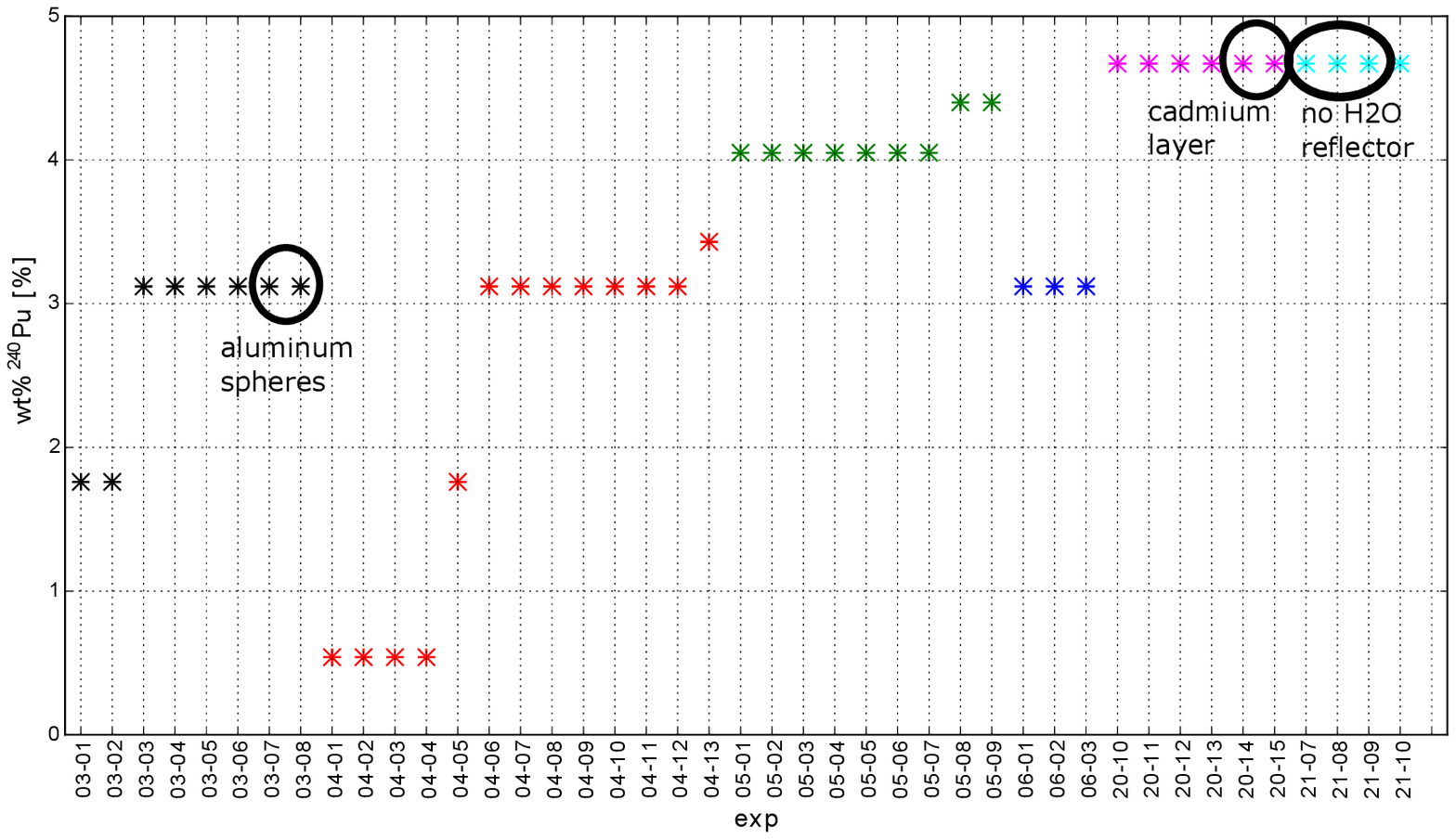}
    \caption{\wtp{240}Pu of all considered cases.}
    \label{fig:weight_pc240Pu}
\end{figure}
Figure \ref{fig:weight_pc240Pu} shows the \wtp{240}Pu for all experiments, for different series in different colors. 
Also the two cases with aluminum sphere, the two cases with additional cadmium layers on the outside of the sphere and the three experiments without water reflector are highlighted. 
The plutonium nuclide vectors for all experiments are given in table \ref{tab:plutoniumVector}.

\begin{table}[ht]
	\centering
		\caption{Plutonium nuclide vector for the experimental series.}
		\label{tab:plutoniumVector}
		\begin{tabular}{|l|l|l|l|l|l|l|}
			\hline
			Experimental 				& \wtp{238}Pu & \wtp{239}Pu & \wtp{240}Pu & \wtp{241}Pu	& \wtp{242}Pu \\
			series       				&  						& 						& 						&							& 				\\
			\hline
			PST-03, 04, 05, 06	& - 					& remainder & see fig. \ref{fig:weight_pc240Pu} &  - & - \\
			\hline                 
			PST-20							& 0.006 			& remainder & see fig. \ref{fig:weight_pc240Pu} & 0.311 & 0.009 \\
			\hline                 
			PST-21							& 0.006 			& remainder & see fig. \ref{fig:weight_pc240Pu} & 0.283 & 0.009 \\
			\hline                 
		\end{tabular}
\end{table}

For series PST-03, 04, 05, and 06 the given temperature is 300$^\circ$ K, for series PST-20 and 21 it is 298$^\circ$ K, following the experimental description. 
Accordingly, the density of the reflecting water is slightly different: 0.9965~g/cm$^3$ and 0.9970~g/cm$^3$.
\begin{figure}[ht]
    \centering
    \includegraphics[width=0.9\textwidth]{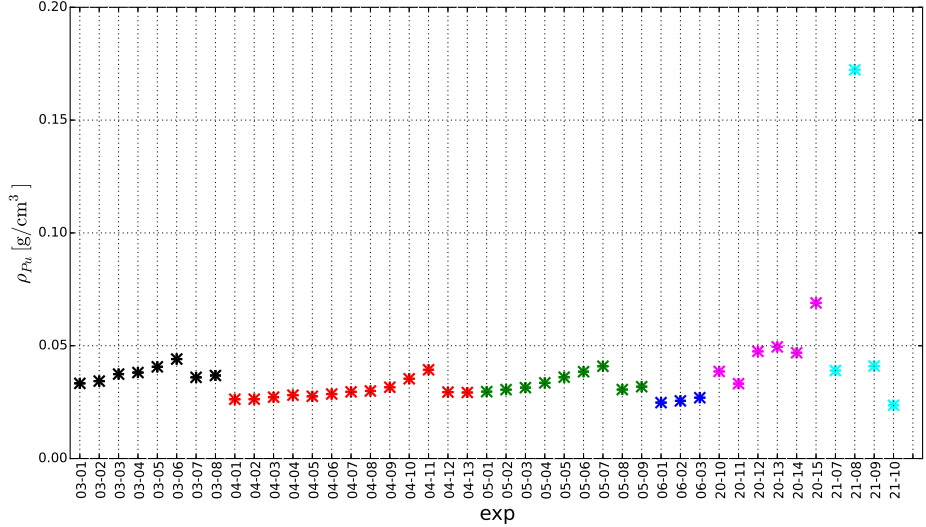}
    \caption{Plutonium density \rhoPu{} of all considered cases.}
    \label{fig:dens_Pu}
\end{figure}
The plutonium densities \rhoPu{} in [g/cm$^3$] vary for all experiments from around 0.025 to 0.07 due to the varying concentration of plutonium nitrate in water, shown in figure \ref{fig:dens_Pu}.
With increasing \rhoPu{} the moderator-to-fuel ratio decreases, leading to a harder spectrum. 
This reduced moderation increases the energy of average lethargy causing fission (EALF). 
\begin{figure}[ht]
    \centering
    \includegraphics[width=0.9\textwidth]{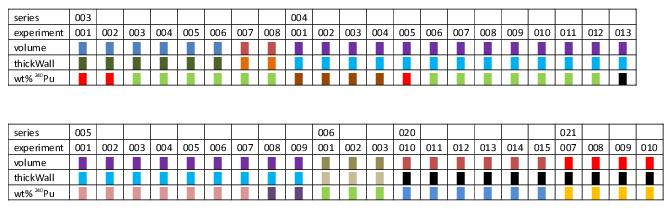}
    \caption{Correlation of varied system parameters. The colors indicate, for which experiments the corresponding parameter is mutually varied. Identical color means identical model parameters.}
    \label{fig:correlationsInput}
\end{figure}

The experiments are correlated, since they share certain system parameters, which are afflicted by experimental uncertainty. 
In the geometrically rather simple experiments presented in this work, these shared system parameters are the volume of the sphere, the thickness of the wall and the \wtp{240}Pu.
It is assumed that experiments with the same values are correlated via these parameters.
Since the experimental descriptions do not describe how the solutions are mixed from their individual components, all used densities (\rhoPu{}, $\rho_{NO_3}$, $\rho_{Fe}$, $\rho_{total}$) are assumed to be independent for each experiment.
The matrix in figure \ref{fig:correlationsInput} shows the modeling assumptions for the correlations of varied system parameters. 
Thereby the parameters are varied mutually for all experiments with the same color box.

For the determination of the correlation coefficients of \keff{} values resulting from shared components in different experiments, we applied a full Monte Carlo method. 
Therein, the variation of system parameters is performed by calculating many samples of the same experiment in which all uncertain parameters are varied simultaneously according to prescribed distribution functions. 
For this method, the GRS tool SUnCISTT (Sensitivities and Uncertainties in Criticality Inventory and Source Term Tool \cite{Behler.2014}) was utilized. 
SUnCISTT controls and organizes the sensitivity and uncertainty analysis for each experiment and determines correlations and covariances between experiments. 

We choose 250 samples for each experiment and normal distribution functions for all parameters. 
The uncertain experimental parameters are listed in table \ref{tab:UncExpParams} with the standard deviation of the distribution for each series. 
\begin{table}[ht]
	\centering
		\caption{Uncertain experimental parameters.}
		\label{tab:UncExpParams}
		\begin{tabular}{|l|l|l|l|}
			\hline
			Uncertain 		& Variable 	& \multicolumn{2}{c|}{Uncertainties} \\
			\cline{3-4}
			experimental 	&  		 			& series & series 				\\
			parameters		 	&						& 03, 04, 05, 06 & 20, 21						\\
			\hline
			\hline
			Total density	& $\rho_{total}$ & 0.03\,\% & 0.4\,\%	\\
			\hline                 
			Pu density 		& \rhoPu{}		 & 1.0\,\% & 1.0\,\% 	\\
			\hline
			Fe density in solution & $\rho_{Fe}$ & 1.4\,\% & - \\
			\hline
			Nitrate density & $\rho_{NO_3}$ & 0.6\,\% & - \\
			\hline
			Acid molarity	& $N_a$ & - & 1.0\,\% \\
			\hline
			Weight\,\%\,$^{238}$Pu  & \wtp{238}Pu & - & 16.67\,\% \\
			\hline
			Weight\,\%\,$^{240}$Pu  & \wtp{240}Pu & 7.0\,\% & 0.75\,\% \\
			\hline
			Weight\,\%\,$^{241}$Pu  & \wtp{241}Pu & - & 1.93\,\% \\
			\hline
			Weight\,\%\,$^{242}$Pu  & \wtp{242}Pu & - & 11.11\,\% \\
			\hline
			Volume & V & 0.3\,\% & 0.25\,\% \\
			\hline
			Wall thickness & $r_{wall}$ & 10.0\,\% & - \\
			\hline
			Temperature H$_2$O & $T_{H_2O}$ & 0.09\,\% & 0.09\,\% \\
			\hline			
		\end{tabular}
\end{table}

The amount of $^{239}$Pu (filled up to 100\,\% Pu), the number densities of the solution and the sphere radii (Pu nitrate, metal, water) for the calculation input are deduced for each sample.

For the generation of the randomly varied parameters a SUnCISTT add-on was used, which was developed in the course of this work and the works presented in references \cite{Peters.2015b, Peters.2016, Peters2016355, Peters.2015}.
The add-on allows to create the random numbers using a configuration file for each experiment.
In this file the following values have to be specified:
the sample size, the type of separator in the list, for each uncertain parameter the following values: variable name, type of distribution (gaussian, normal, $\beta$, linear-, logarithmic-, or exponential-increase), a seed for the random number generator, details of the distribution (mean and standard deviation or limits of the distribution, type of $\beta$ distribution or base for logarithm and exponent), number of instances, a comment. 
The used random number generator is based on a deterministic method using the Mersenne Twister sequence to generate pseudo-random numbers.
Thus, by setting the seed equal in two configuration files, the generated random numbers are the same. 
Accordingly the experiments are correlated via this parameter. 
For all independent variables, no seeds are specified, so that the random numbers are independent for each experiment. 
Following the pattern of correlated parameters shown in figure \ref{fig:correlationsInput}, the same seeds are preset for equally colored boxes. 

The mean values of the derived number density distributions were compared to the nominal values given in the experimental description as a check of plausibility of the calculations (deviation $<$ 0.05\,\%).

\section{Results}
\label{sec:results}

In this section the results of the calculations are presented containing the \keff{} values, the sensitivities of \keff{} on the varied parameters, the resulting correlations of \keff{}, the correlations due to nuclear uncertainties and a comparison to the DICE data bank.

\subsection{\keff{} values}
\label{secSub:keff}

\begin{figure}[ht]
    \centering
    \includegraphics[width=0.9\textwidth]{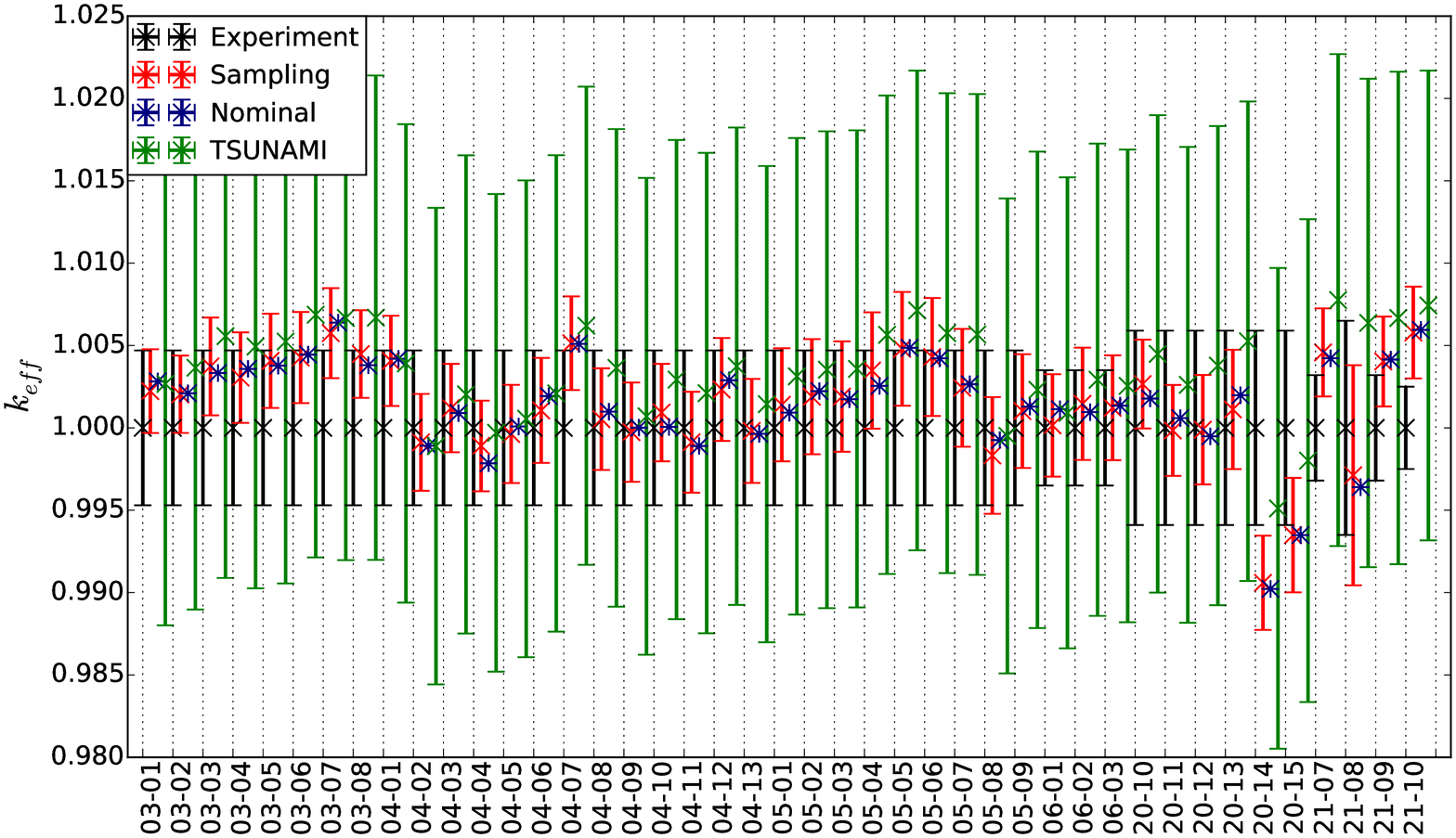}
    \caption{Range of the calculated \keff{} values.}
    \label{fig:keffCalc}
\end{figure}

Figure \ref{fig:keffCalc} shows the calculated and experimental \keff{} values for all 43 experiments and the corresponding  1$\sigma$ deviation. 
The black crosses show the experimental values \keff{exp}$=1$, the error bars represent the experimental uncertainties due to the uncertainties of system parameters (see table \ref{tab:experiments}), which were combined by the evaluators via error propagation.
The blue crosses indicate the nominal calculations.
Also shown are the mean values and standard deviations of the sampling calculations due to the variation of system parameters (red, see section \ref{secSub:corrSystemParams}) and of nuclear data (green, TSUNAMI, see section \ref{secSub:corrNuclearUnc}).

Compared to the other experiments the two experiments with cadmium coatings of the spheres (PST-20-14 and 15) deviate significantly towards lower values ($\Delta \keff{} \approx{}  -0.0112$, bzw. $-0.0080$).
Cadmium is a strong neutron absorber in the thermal range, so that effectively the impact of the water reflector is diminished. 
The thickness of 0.03 inch (0.762 mm) used here reduces the thermal neutron flux roughly by a factor of 6\ep{3} (calculated from a typical neutron reduction by a factor of 1\ep{5} per 1 mm cadmium, deduced from a total macroscopic cross section of 115 cm$^{-1}$ \cite{cadmium}). 
Therefore these two cases where recalculated by assuming three different thicknesses of cadmium of 0.02 inch (0.508 mm), 0.01 inch (0.254 mm) and 0 inch in order to check, if this variation is the reason for the deviation from the other calculations. 
The value of 0.02 inch was typical for similar experiments, according to the experimental description.
Table \ref{tab:keffCadmium} summarizes these calculations.

\begin{table}[ht]
	\centering
		\caption{\keff{} values from the variation of the cadmium thickness.}
		\label{tab:keffCadmium}
		\begin{tabular}{|l|l|l|l|l|}
			\hline
			Experiment 	& \keff{\mathrm{0.03\ inch}} & \keff{\mathrm{0.02\ inch}} & \keff{\mathrm{0.01\ inch}} & \keff{\mathrm{0.0\ inch}} \\
			\hline
			\hline
			PST-020-14	& 0.9902 & 0.9912 & 0.9920 & 1.0673 \\
			\hline                 
			PST-020-15	& 0.9935 & 0.9946 & 0.9949 & 1.0751 \\
			\hline                 
		\end{tabular}
\end{table}

One sees, that the reduction from 0.03 to 0.02 or even only 0.01 inch does not significantly increase \keff{} to values comparable to the remaining calculations, so that even a hypothetical gross error in the measurement of the thickness cannot resolve the deviation. 
On the other hand, the \keff{} increases approximately 7\,\% to values of 1.07 if no cadmium is present. 
Our working hypothesis is, that KENO-Va overestimates the influence of cadmium, possibly due to nuclear cross sections. 
To determine a systematic effect of cadmium on the validation of criticality codes, further studies have to be performed. 
The analysis of these two cases shows, that one has to take special care, when validating codes with systems containing cadmium. 
These two experiments are used in the following with the original assumed 0.03 inch of cadmium coating.

In order to examine the influence of the sphere material on criticality, all calculations were repeated assuming vacuum instead of stainless steal or aluminum. 
By that the average \keff{} increases by about $\Delta\keff{} = 0.0099$ compared to the calculation including the spheres (not considering cases with aluminum spheres, cadmium coating and without sphere). 
This shows the neutron absorbing effect of steal, which is governed by the contained iron with an average thermal absorption cross section of around 2.7 barn. 
On the other hand, the cases with aluminum spheres and cadmium layers on the sphere show a small decrease of \keff{} with removal of the spheres ($\Delta\keff{Al} = -0.0018 $ and $\Delta\keff{Cd} = -0.0048$). 
This can be explained by the also present reflecting effect of the sphere material. 
Since the absorbing effect of steal is a lot smaller than the one of cadmium, the increase in these cases is much smaller. 
Aluminum on the other side has a much smaller cross section than steal, so that the much smaller change in \keff{} is explicable. 

To identify dependencies between \keff{} and other physical parameters, a trend analysis was performed against \rhoPu{}, the \wtp{240}Pu, $\rho_{NO_3}$, EALF and the ratio of moderator to fuel $H/Pu_{total}$.
No relevant and significant dependencies of \keff{} on these parameters could be identified.

\subsection{Uncertainty analysis}
\label{secSub:uncAnalysis}

This section presents the results of an uncertainty analysis performed to deduce to what extent experimental uncertainties influence the calculated \keff{} value and which are the leading effects.

The calculated \keff{} values and their standard deviations are included in figure \ref{fig:keffCalc} in red. 
The mean values agree very well with the nominal values. 
The standard deviation is comparable to the experimental uncertainty, but for series PST-03, 04, 05 and 20 it is 30\,-\,50\,\% smaller. 
This can be attributed to the fact, that for these series, the given experimental uncertainty is not calculated directly, but assumed from calculations of other experiments. 
For series PST-06 and 21 we find an excellent agreement of the standard deviations and experimental uncertainties.

\begin{figure}[!ht]
    \centering
    \includegraphics[width=1.0\textwidth]{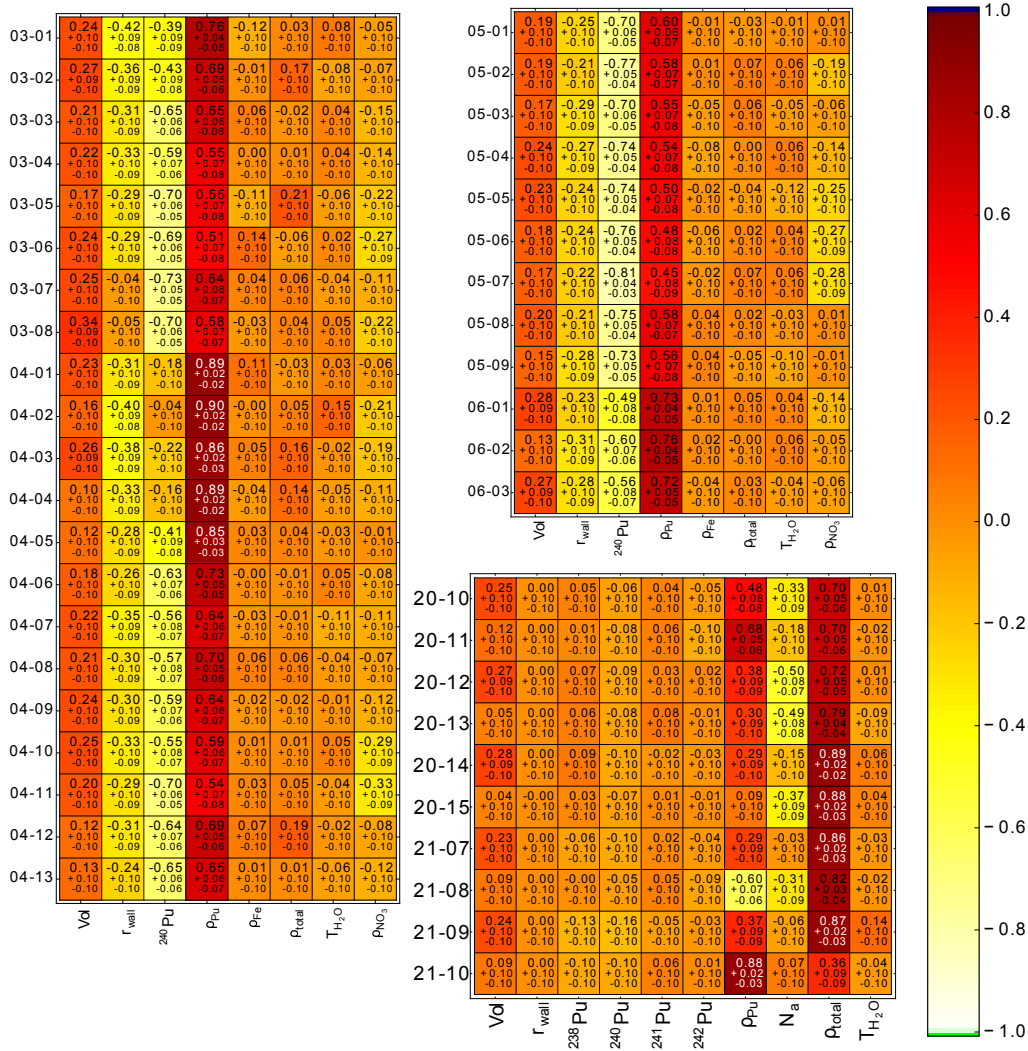}
    \caption{Sensitivities of \keff{} on uncertain parameters for all analyzed experiments. Numerical values shown for mean and the 95\% confidence level.}
    \label{fig:sensitivities}
\end{figure}

For all experiments the Pearson correlation coefficients were calculated between the varied parameters and \keff{}, shown in figure \ref{fig:sensitivities}.
This gives a measure for the influence of the variation of each uncertain parameter on the uncertainty of \keff{} and can demonstrate the leading effects. 
Note that this is not the sensitivity of \keff{} on the uncertain parameters, but the sensitivity of \keff{} on the actual variation of the uncertain parameters.

For the first four series (PST-03, to -06), the two leading effects are a negative correlation between \keff{} and \wtp{240}Pu, and a positive correlation with the Pu density \rhoPu{}.
The first relation can be explained by the reduction of the most reactive plutonium isotope $^{239}$Pu. 
The second by an increase of plutonium atoms available for fission.
The \keff{}-decreasing effect of the small decrease of $H/Pu$ by the increase of \rhoPu{} in these under-moderated systems can be neglected.
A small positive correlation exists to the sphere volume $V$. 

All these correlations are obvious: The more material is present, the higher is \keff{}.
Also notably is, that the sensitivity to \wtp{240}Pu increases with \wtp{240}Pu itself since the absolute variation increases.  
The small negative correlation with the wall thickness $r_{wall}$ can be understood by the neutron absorption of stainless steal. 
Therefore its value is not significantly different from zero for the two experiments with aluminum sphere PST-03-07 and 08. 
The uncertainty of the density of the impurity iron $\rho_{Fe}$, of the total density $\rho_{total}$ and of the temperature $T_{H_2O}$ have almost no significant effect on \keff{}. 
For the total solution density $\rho_{total}$ this is certainly attributed to the very small given uncertainty of only 0.03\,\% for these experiments.

For the second set of experiments (PST-20 and -21) the situation is slightly different. 
Here the leading effect is the total solution density $\rho_{total}$, which has a 13 times higher uncertainty of 0.4\,\% leading to a strong positive correlation. 
Additionally the given uncertainty of \wtp{240}Pu in this second set is a factor of 100 smaller than in the first set, so that its influence on \keff{} disappears almost completely.
The next effect is a negative correlation with the acid molarity $N_a$.
An increase of $N_a$ leads to an increase of $\rho_{NO_3}$ and a decrease of $\rho_{H_2O}$. 
$\rho_{NO_3}$ increases the number density of $^{14}$N having a considerable neutron absorbing effect, $\rho_{H_2O}$ drives the moderation ratio away from its optimum value, both explaining the negative impact on \keff{}. 
The mostly positive correlation with \rhoPu{} is evident due to the same effect as for the first set of experiments. 


\subsection{Correlations due to system parameter uncertainties}
\label{secSub:corrSystemParams}

\begin{figure}[!ht]
    \centering
    \includegraphics[width=1.0\textwidth]{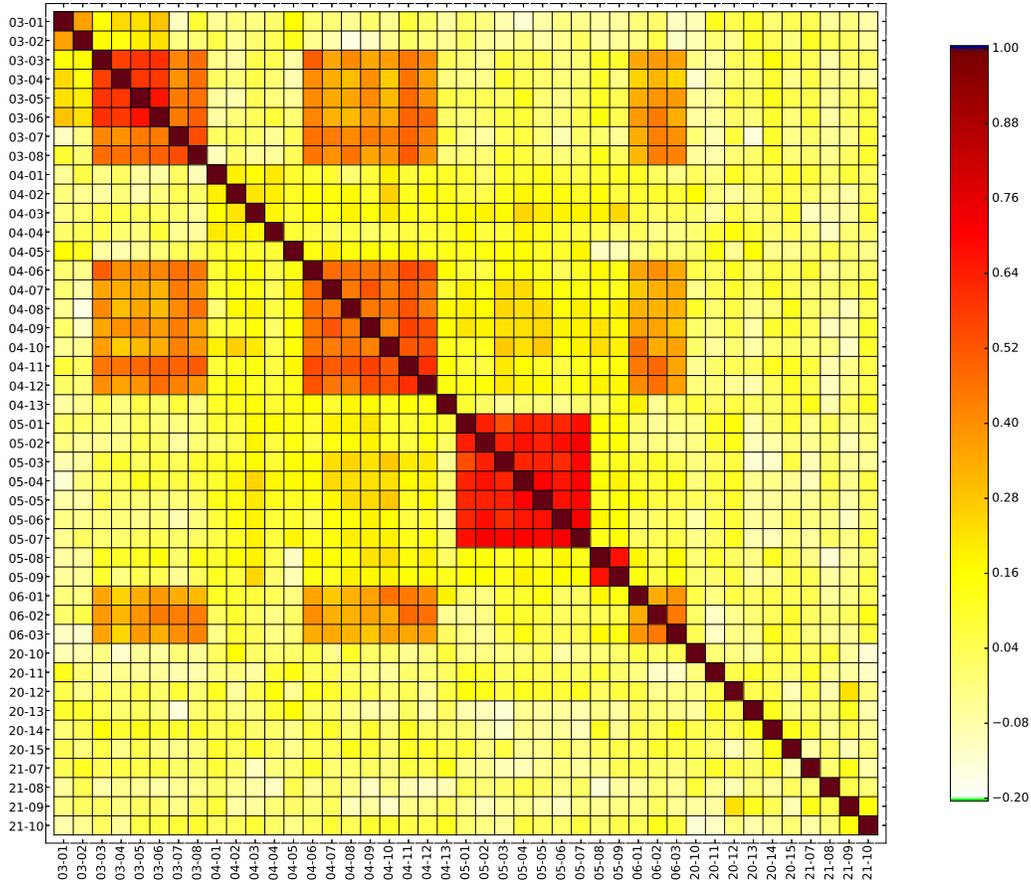}
    \caption{Correlation matrix of \keff{} between all experiments due to the partly mutual variation of uncertain system parameters.}
    \label{fig:correlations}
\end{figure}

In this section the calculated correlation coefficients \corr{} between the \keff{} values of all analyzed experiments are discussed.
The different experiments are correlated due to shared system parameters (figure \ref{fig:correlationsInput}). 
In general, the correlation values range from slightly negative values to 0.7, but most correlation coefficients are in the range of $[-0.1,0.3]$.
Since the error of \corr{} is in the range of 0.1 for values around $\corr{} = 0$ for the used 250 samples \cite{Peters2016355}, most correlations between experiments can be considered statistically not or only slightly significant.
Higher correlation coefficients can be found within the experimental series PST-03 (experiments 03 to 08), PST-04 (experiments 06 to 12), PST-05 (experiments 01 to 07 and 08, 09) and PST-06 (experiments 01 to 03). 
Further, three blocks of higher correlation coefficients between experimental series can be identified: 
Experiments PST-03-03 to -08 with PST-04-06 to -12 and experiments PST-06-01 to -03, and PST-04-06 to -12 with PST-06-01 to -03. 

All cases of series PST-20 and 21, PST-04-01 to -05 and -13 are uncorrelated to the others. 
This can be explained, since the sensitivities of \keff{} on the mutual varied parameters are much smaller than the ones on the individually varied parameter.

Comparing the results shown in figure \ref{fig:correlations} with figure \ref{fig:correlationsInput} one can see, that almost all blocks of higher correlations are due to the same \wtp{240}Pu.
This is the only one of the two leading effects of the sensitivity of \keff{}, which is varied mutually: PST-03-01 and -02, PST-03-03 to -08, PST-04-06 to -12, PST-05-01 to -07, PST-05-08 and -09, PST-06-01 to -03, and PST-03-03 to PST-03-08 with PST-04-06 to -12 and with PST-06-01 to -03.

\subsection{Correlations due to nuclear data uncertainties}
\label{secSub:corrNuclearUnc}

\begin{figure}[!ht]
    \centering
    \includegraphics[width=.9\textwidth]{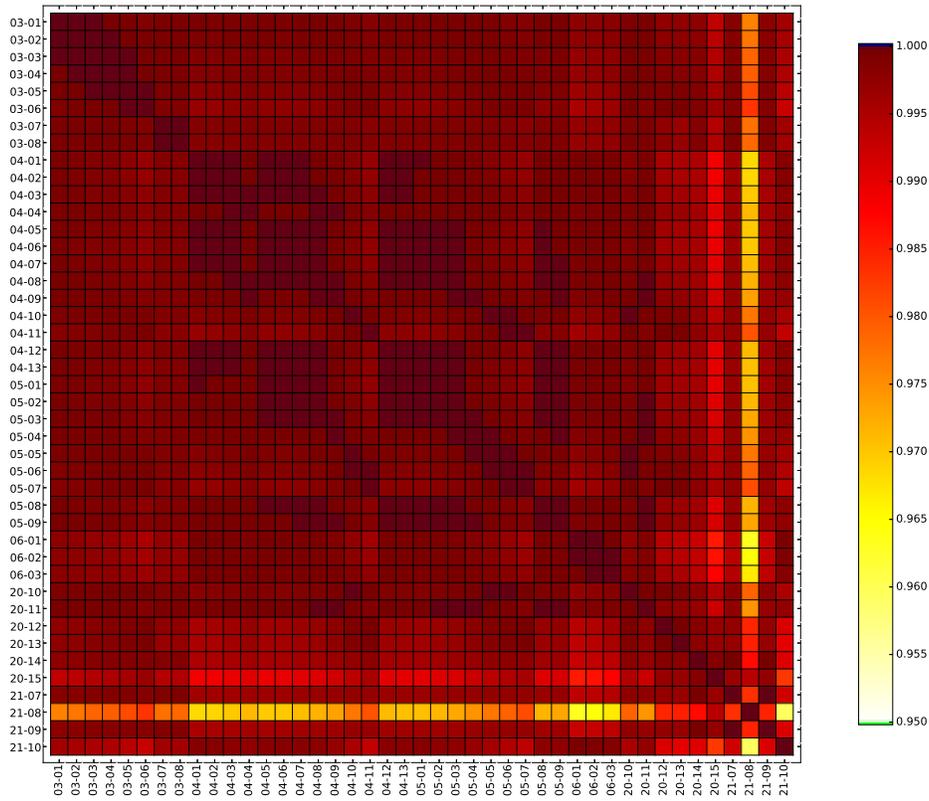}
    \caption{Correlation matrix of \keff{} between all experiments due to uncertainties in the nuclear data. All values are above 0.95. Note the different color scale w.r.t. figure \ref{fig:correlations}.}
    \label{fig:correlationsNuclData}
\end{figure}

The correlations between experiments due to uncertainties in the nuclear data were examined.
While the correlations due to system parameters are strongly depending on the interpretation of the given experimental data, the uncertainties of the involved nuclear processes mainly depend on the material composition and the choice of the cross section library including its covariance matrix.
They can be assumed independent from system parameter uncertainties.
We used the sequence TSUNAMI-3D-5K to calculate the sensitivities of \keff{} on the nuclear processes.
TSUNAMI-IP was used to infer the correlation values \ck{} between experiments due to these processes.

The \keff{} values and their uncertainties due to nuclear data are added to figure \ref{fig:keffCalc} in green.
The nominal values of the TSUNAMI calculations are higher than the calculations with CSAS5.
One possible explanation is the use of different cross section libraries (continuous energy in CSAS5 and 238-group structure in TSUNAMI).
Their uncertainties due to nuclear data uncertainties are clearly dominant over the uncertainties due to system parameters.
This was already found for the LEU-COMP-THERM experiments in references \cite{Peters.2015b, Peters.2016, Peters2016355, Peters.2015}.

The \ck{} values between all experiments except PST-21-08 have values above 0.98 and can be assumed to be identical in terms of sensitivity to nuclear data. 
Due to its different moderation ratio, experiment PST-21-08 has slightly lower correlations with the other experiments down to 0.96, but still shows very high values.

\subsection{Comparison with DICE}
\label{secSub:dice}
Our results can be compared to available data of the "Database for the International Criticality Safety Benchmark Evaluation Project" (DICE \cite {DICE}). DICE is still in the development phase and also subject to possible data entry errors and omissions. However, it provides some information on correlations on a vast number of experimental series described in the ICSBEP Handbook \cite{NuclearScienceCommittee.September2014}. 
The relevant data for our analysis is shown in table \ref{tab:DICEcorr}. 
There is no data available neither for the correlation of PST-21 nor for results on the case level details.
A "+"-sign indicates strong correlations between experiments when one or several uncertain benchmark parameters are correlated, which are major contributors to the overall benchmark \keff{} uncertainty.
A "(+)"-symbol indicates a 100\% correlation. 

The currently limited data in DICE does only partly agree with the findings of our more detailed analysis (figure \ref{fig:correlations}).
On one side, we also found statistically significant correlation coefficients between series for experiments PST-03-03 to -08 with PST-04-06 to -12 and PST-06.
However all these cases have correlation coefficients below 0.6.

On the other side, all experiments of PST-20, PST-04-01 to -05 and -13 show no statistically significant correlation coefficients with any other investigated experiment.
 
\begin{table}[ht]
	\centering
		\begin{tabular}{|l|c|c|c|c|c|}
			\hline
			& PST-03 & PST-04 & PST-05 & PST-06 & PST-20 \\
			\hline
			\hline
			PST-03 & (+)  & + & + &+  & +  \\
			\hline                 
			PST-04 & + & (+) & +  & + & +  \\
			\hline                
			PST-05 &  +& + & (+) & + &  +\\
			\hline                
			PST-06 & + & + & + & (+) &  + \\
			\hline                
			PST-20 & + & + & + &  + & (+)  \\
			\hline                
		\end{tabular}
	\caption{Information on possible correlations due to shared experimental components taken from \cite{DICE}. A "+"-sign indicates strong correlations, a "(+)"-symbol 100\% correlation.}
	\label{tab:DICEcorr}
\end{table}

\begin{table}[ht]
	\centering
		\begin{tabular}{|l|c|c|c|c|}
			\hline
			& PST-03 & PST-04 & PST-05 & PST-06 \\
			\hline
			\hline
			PST-03 & 999  & 997 & 998 & 995  \\
			\hline                 
			PST-04 & 997 & 999 & 998  & 998 \\
			\hline                
			PST-05 & 998 & 998 & 999 &  997 \\
			\hline                
			PST-06 & 995 & 998 & 997 & 1000 \\
			\hline                
		\end{tabular}
	\caption{Evaluation level correlation due to nuclear data taken from \cite{DICE}. The numeric values lie between -1000 and 1000 and are averages of the case level correlations between the evaluations.}
	\label{tab:DICEck}
\end{table}

Table \ref{tab:DICEck} shows the values corresponding to the E-metric in TSUNAMI, which is the dot product between sensitivity vectors. The values are the mean values taken over all case level detail values and taken directly from \cite{DICE}. There is no weighting by nuclear data covariances because the intent is to show how similar the sensitivities are between cases. A value of 1000 indicates identity and would be comparable to a correlation coefficient of 1 in our figure \ref{fig:correlationsNuclData}. Our \ck{} values are in very good agreement with the data given in DICE.

%
%
\section{Discussion}
\label{sec:discussionConclusion}

We investigated correlations of \keff{} values of a set of 43 critical experiments of plutonium nitrate spheres in aqueous solution.
The experiments were conducted in 6 series and are described as PU-SOL-THERM series number 03, 04, 05, 06, 20, and 21 in the ICSBEP handbook \cite{NuclearScienceCommittee.September2014}.
To derive correlation coefficients due to experimental uncertainties we performed a full Monte-Carlo analysis using SUnCISTT \cite{Behler.2014} to steer the CSAS5 sequence of SCALE 6.1.2 \cite{ORNL.2012}.
The correlation coefficients due to nuclear data uncertainties were calculated using TSUNAMI of SCLAE 6.1.2. 
The latter are in concordance with values given in \cite{DICE}.

The assumptions made to model the experimental data and to calculate the \keff{} values are described in detail in section \ref{sec:modelingAssumptions}. 
All calculated \keff{} results are within 1$\sigma$ agreement with the experimental values except for two cases (figure \ref{fig:keffCalc}). 
Our \keff{} results for the experiments with cadmium coatings of the sphere  (PST-20-14 and 15) deviate from the experimental value. 
We found, that the reduction from 0.03 to 0.01 inch of cadmium coating does not significantly increase \keff{} to values comparable to the experimental values.
On the other hand, \keff{} increases by approximately 7\,\% to values of 1.07 if no cadmium is present. 
Our analysis overestimates the influence of cadmium, possibly due to the cadmium nuclear cross sections in the used continuous energy library. 

We determined the correlation coefficients due to shared uncertain modeling parameters (figure \ref{fig:correlationsInput}) and presented them color coded in figure \ref{fig:correlations}. 
The correlation coefficients are found to be low and only for some experiments (PST-05-01 to -07) reaching values around $\corr{} \approx 0.7$. 
All other experiments have smaller correlation coefficients and most of them statistically not significant. 

The only variation introducing a correlation effect between different series is the identical \wtp{240}Pu of the experiments PST-03-03 to -08, PST-04-06 to -12, and PST-06.
These results support on the one hand the data provided by DICE \cite{DICE} for possible correlations of experiments in the series PST-03 to -06. 
On the other hand, we found, that for these experiments, the correlation due to shared experimental components are small and are likely to have a negligible effect in validation procedures. 
We find further, that certain experiments (PST-03-01 and -02, PST-04-01 to -05 and -13, series PST-20 and -21) have no statistical significant correlation coefficient with experiments from any other series. 
The shared components for these experiments do not introduce correlations due to their small given uncertainty in the experimental description. 
This is in contrast to e.g. our findings for lattices of fuel rods, where shared experimental components can introduce large correlation coefficients in the data \cite{Peters.2015b, Peters2016355, Peters.2015} due to higher sensitivities of \keff{} to these shared components and their corresponding uncertainties.

\section{Conclusion}
\label{sec:Conclusion}

To determine a possible systematic effect of cadmium on the validation of criticality codes further work has to be done.
We found that our analysis overestimates the influence of cadmium in PST-20-14 and -15, possibly due to the cadmium nuclear cross sections.
However, one has to take special care, when validating codes with systems containing cadmium.

We found for our calculations that the presence of shared components within an experimental series does not necessarily lead to statistically significant values of correlation coefficients.
If the shared experimental components are very well known in the sense of comparable small uncertainties or if these components play only a minor role on $\Delta\keff{}$ (determined e.g. by means of a sensitivity analysis), the effect on the correlation coefficient is negligible.
We conclude, that for the determination and understanding of correlation coefficients it seems to be inevitable to perform a detailed sensitivity analysis of the underlying modeling assumptions and a thorough review, analysis and interpretation of the experimental description.

\section*{Acknowledgments}
The work was funded by the German Federal Ministry for the Environment, Nature Conservation, Building, and Nuclear Safety, and supported by the German Federal Office for Radiation Protection.
We thank Tatiana Ivanova and Ian Hill for their help in using DICE.


\end{document}